\documentclass[prl,aps,twocolumn,amsmath,amssymb,showpacs]{revtex4}
\usepackage{graphicx}
\usepackage{bm}
\begin{document}

\title{Single and two-particle motion of heavy
particles in turbulence}

\author{Itzhak Fouxon$^1$}
\author{P\'eter Horvai$^2$}

\affiliation{$^1$ Racah Institute of Physics,
  Hebrew University of Jerusalem,
  Jerusalem 91904,
  Israel}
\affiliation{$^2$ Mathematics Institute, University of Warwick,
Coventry CV4 7AL, UK}

\begin{abstract}

We study motion of small particles in turbulence when the particle
relaxation time falls in the range of inertial time-scales of the
flow. Due to inertia, particles drift relative to the fluid. We show
that the drift velocity
is close to the Lagrangian velocity increments of turbulence at the
particle relaxation time. We demonstrate that the collective drift
of two close particles makes them see local velocity increments
fluctuate fast and we introduce the corresponding Langevin description for
separation dynamics.
This allows to describe the behavior of 
the Lyapunov exponent and give the analogue of Richardson's law for
separation above viscous
scale.

\end{abstract}


\maketitle

Motion of small particles in a fluid, due to random molecular
forces, is the subject of the classical theory of Brownian motion.
Scale separation between the particle relaxation time and
time-scales of the forces allows to introduce an effective Langevin
description of the driving force as white noise in time \cite{Ma}.
In contrast, here we consider the situation where the random driving
force originates not from the microscopic motions, but rather from the
macroscopic turbulent motion of the surrounding fluid \cite{Frisch}.
The limit of the particle relaxation time much larger than the
characteristic time-scales of turbulence (very heavy particles) can
be described as in the Brownian motion case
\cite{BecCenciniHillerbrand}. In the opposite limit, when particle
relaxation time is much smaller than the characteristic time-scales
of turbulence (very large friction), particles follow the flow
closely and the two-particle dispersion -- of interest to us here,
-- is approximately the same as for fluid particles.  In this Letter
we study the intermediate case of heavy particles, where the
relaxation time falls in the range of flow time-scales corresponding
to the inertial interval of turbulence. This precludes Langevin
description for the single-particle motion. However, for {\em two}
particles, because of their collective drift relative to the fluid,
velocity increments determining the separation do vary fast. This
allows to introduce effective Langevin description for the dynamics
of separations. The description enables us to find several results
on particle behavior in turbulence valid beyond Kolmogorov theory.

Behavior of small inertial particles in turbulence has received much
attention lately
\cite{Shaw,Nature,Piterbarg,engineering,MWB,DFTT,Horvai,BBBCC,BBB,BLG,
Grisha,MaxRil,Maxey,FH}.  This problem has many
applications including rain formation in clouds \cite{Shaw,Nature},
ocean physics \cite{Piterbarg} and engineering \cite{engineering}.
Theoretical study of the problem mostly involved modeling turbulence
by a white noise in time, Gaussian velocity field, known as
Kraichnan model.
Even in that case theoretical study is rather difficult, 
analytic results were mainly obtained for the one-dimensional case
\cite{MWB,DFTT}. The limit of heavy particles, considered here for
turbulence, was studied numerically for Kraichnan model in
\cite{BecCenciniHillerbrand,Horvai}. For turbulence, numerical
studies of intermediate regime of moderately heavy particles were
performed in \cite{BBBCC,BBB,BLG}.

We consider the motion of a small spherical particle
in an incompressible, statistically steady, turbulent flow ${\bm
u}({\bm x}, t)$. We assume that the drag force acting on the
particle obeys Stokes' law. Designating the particle position and
velocity by ${\bm x}(t)$ and ${\bm v}(t)$, Newton's law reads
%
%
%
\begin{equation}
\label{Newton}
  \dot{\bm x}
=
  \bm v
\,\,,\qquad
  \dot{\bm v}
=
  -\frac{\bm v - \bm u(\bm x(t),t)}{\tau}
\,.
\end{equation}
Here $\tau = (2/9)(\rho_0/\rho)(a^2/\nu)$, where $\rho_0$ and $a$ are the
particle density and radius, while $\rho$ and $\nu$ are the fluid's density and
kinematic viscosity. We briefly review relevant properties of $\bm
u(\bm x, t)$, see e.g.\ \cite{Frisch} for details. Velocity field,
excited at the integral scale $L$, fluctuates in a wide (inertial) range
of spatial scales $\eta\ll l\ll L$. The characteristic velocity
$u_l$ of fluctuations at a scale $l$ is related to the temporal scale
$t_l$ by $u_lt_l/l\sim 1$. At the viscous scale $\eta$ we have
$t_{\eta}\sim\eta^2/\nu$. For moderate Reynolds numbers one can use
Kolmogorov theory (below K41) that gives $u_l\sim (\epsilon
l)^{1/3}$, $t_l\sim \epsilon^{-1/3}l^{2/3}$ and
$\eta\sim(\nu^3/\epsilon)^{1/4}$, where $\epsilon$ is the mean
energy injection rate. Eq.~(\ref{Newton}) is valid if $\eta\gg a$
and inertia-induced particle drift relative to the flow, described
by ${\bm w}(t) \equiv {\bm v}(t)-{\bm u}({\bm x}(t), t)$, has small
Reynolds number $w a/\nu$ \cite{MaxRil}. We shall consider $t_{\eta}\ll
\tau\ll t_L$, which implies that particles are heavy, $\rho_0/\rho\sim
(\eta^2/a^2)(\tau/t_{\eta})\gg 1$, justifying the neglect of such effects
as added mass in Eq.~\eqref{Newton} \cite{MaxRil}. Beyond K41,
quantities like $\eta$ and $w$ have strong
spatiotemporal fluctuations, and in that setup we will refer to
their local (in space and time) values on statistically relevant
events.

We first consider the drift velocity $\bm w$. From Eq.~(\ref{Newton}), in the steady state,
${\bm w}=\int_{-\infty}^0 \delta_t^P \bm u\exp(t/\tau)dt/\tau$,
where $\delta_t^P \bm u \equiv {\bm u}({\bm x}(t), t)-{\bm u}(\bm 0,
0)$ is the turbulent velocity difference in particle frame and we
chose $t=0$ and $\bm x(0)=0$.  The difference $\delta_t^P \bm u$
is analogous to the Lagrangian difference $\delta_t^L \bm u \equiv
{\bm u}({\bm q}(t), t)-{\bm u}(\bm 0, 0)$ where ${\bm q}(t)$ is a
fluid particle trajectory obeying ${\dot {\bm q}}={\bm u}({\bm
q}(t), t)$ and $\bm q(0)=\bm 0$.  Just as $\delta_t^L u$, the
increment $\delta_t^P u$ should be, on a rough scale, a
non-decreasing function of $|t|$, growing at most as a power law.
Then the integral for $\bm w$ yields the order of magnitude estimate
$w \sim \delta_{\tau}^P u$ and we also see that the characteristic
time of variations of $\bm w$ is $\tau$. Let us show that from $w
\sim \delta_{\tau}^P u$ one can pass to $w \sim \delta_{\tau}^L u$,
that is $\delta_{\tau}^P u\sim \delta_{\tau}^L u$. We write the
telescopic sum $\delta_t^P \bm u = \delta_t^L \bm u + \delta \bm v$, where $\delta \bm v \equiv \bm u(\bm
x(t),t) - \bm u(\bm q(t),t)$ is the velocity difference of the separating
fluid and inertial particles. The separation is due to the
combined effect of the inertial drift and the explosive
separation of trajectories in the inertial range \cite{review}, so that
$\delta v\lesssim max[w, \delta_t^L  u ]$, where $\delta_t^L  u $
is the relative velocity of explosively separating particles \cite{Frisch}. 
Thus, at $t$ such that $w\lesssim \delta_t^L u$,
one has $\delta v \lesssim \delta_t^L u$ and, then, from
the telescopic sum, $\delta_t^P u \sim \delta_t^L u$.
This,
combined with $w \sim \delta_\tau^P u$, allows to show $w \sim
\delta_\tau^L u$.  Since $\delta_t^L u$ roughly grows with $t$ at
$0<t<t_L$ and $0=\delta_{t=0}^Lu<w\lesssim \delta_{t_L}^Lu$ (the
latter is just an upper bound for $w$), there exists $t_*$ such that
$w \sim \delta_{t_*}^L u$.  At this $t_*$, from the previous
considerations one has $\delta_{t^*}^P u \sim \delta_{t^*}^L u \sim
w \sim \delta_\tau^P u$.  Finally $\delta_t^P u$ roughly increases
monotonically for timescales less than $L/w \gtrsim L /
\delta_{t_L}^L u \sim t_L$, so we conclude that $t^* \sim \tau$,
whence $w \sim \delta_\tau^L u$. This means that
particles follow only the flow fluctuations with time-scales larger than
$\tau$. Though expectable,
the result uses specific properties of turbulence and does not hold
for any $\bm u(\bm x, t)$, where the relation between the spatial and
temporal fluctuations is different. For the particle velocity, from
$w\ll u$, we have ${\bm v}\approx {\bm u}(\bm x(t), t)$. Note that
$\bm w$ is accessible experimentally through acceleration $\bm
a\equiv {\dot {\bm v}}=-\bm w/\tau$. Also note that $w \sim
\delta_\tau^L u$ applies at any $\tau$: at $\tau\gtrsim t_L$ it is
trivial, while at $\tau\ll t_{\eta}$ one has ${\bm w}\approx
-\tau[\partial_t{\bm u}+({\bm u}\cdot {\bm \nabla}){\bm u}]\approx
\delta_{\tau}^L \bm u$ \cite{Maxey}.

The local equality $w\sim\delta_{\tau}^Lu$ suggests that the
time-averages along the particle trajectory $\bm x(t)$ satisfy
\begin{eqnarray}&&
\tau^n\langle a^n\rangle=\langle w^n\rangle\sim \langle
(\delta_{\tau}^L u)^n \rangle\sim
  u_L^{n} (\tau/t_L)^{\gamma_n}. \label{anomalous}
\end{eqnarray}
The anomalous exponents $\gamma_n$ could differ from their
counterparts for the more usual $\langle (\delta_{\tau}^L u)^n \rangle_q$, where
the averaging is along the fluid particle trajectory
\cite{Frisch,Landau}, -- inertial particles tend to concentrate
preferentially in specific regions of the flow \cite{Shaw}. Yet
the difference seems unimportant: introducing $l_{\tau}$
by $t_{l_{\tau}}\sim \tau$, one has $\langle (\delta_{\tau}^L u)^n
\rangle\sim \langle [\delta u(l_{\tau})]^n \rangle$, where
$\delta \bm u({\bm R})\equiv {\bm u}({\bm x}+{\bm R})-{\bm
u}({\bm x})$, and $\langle [\delta u(l_{\tau})]^n \rangle$ allows additional spatial averaging
over $l_{\tau}$ vicinity of $\bm x(t)$,
beyond which 
the preferential
concentration is expected to be small \cite{BLG}. In K41, similarity of properties
of $w$ and $\delta_{\tau}^Lu$
follows by dimensional analysis: 
one has $w\sim
\sqrt{\epsilon \tau}$ and $\langle w^n \rangle \sim(\epsilon
\tau)^{n/2}$ like for $\delta_{\tau}^Lu$ \cite{Landau}.

We now consider two-particle motion.  We assume the particle separation
${\bm R}={\bm x}'-{\bm x}$ much larger than radius $a$, so that
hydrodynamic interactions between particles are negligible. Then
each particle satisfies Eq.~\eqref{Newton}, producing
\begin{equation}
\label{eq:2p-gen}
  \ddot{\bm R} + \dot {\bm R}/\tau
=
  \delta {\bm u}({\bm R})/\tau
\,.
\end{equation}
At $R\gg l_{\tau}$ particle dispersion, driven by $\delta \bm u(\bm R)\sim u_R$, is determined by turbulent fluctuations slower than $\tau$, and the separation is like for fluid particles: ${\dot {\bm R}}={\bm v}'-{\bm v}=\delta \bm u({\bm R})+{\bm w}'-{\bm w}\approx \delta \bm u({\bm R})$, by $w\sim u_{l_{\tau}}\ll u_R$.
In contrast, at $R\ll l_{\tau}$ dispersion laws peculiar for inertial particles hold
(in K41 $l_{\tau}\sim \epsilon^{1/2}\tau^{3/2}
\sim \eta(\tau/t_{\eta})^{3/2}$).
We first treat $R(t)\ll \eta$
where $\delta u_i(\bm R)\approx R_j\nabla_j u_i(\bm x(t), t)$ and
\begin{equation}
\label{viscousseparation}
  \ddot{\bm R} + \dot {\bm R}/\tau
=
  ({\bm R}\cdot\bm\nabla){\bm u}/\tau
\,.
\end{equation}
The main characteristics of Eq.~\eqref{viscousseparation} is the Lyapunov exponent
$\lambda_1=\lim_{t\to\infty}\ln[p(t)/p(0)]/t=\langle {\dot p}/p\rangle$ describing the exponential growth
of the distance $\bm p\equiv (\bm R, \tau {\dot {\bm R}})$ between two infinitesimally
close trajectories in the phase space of Eq.~\eqref{Newton}.
Here $\lambda_1$ also describes the growth of $\bm R$ and ${\dot
{\bm R}}$ separately, so it can be observed via
spatial trajectories of close particles with small relative
velocity.

We assume that the time-average $\langle {\dot p}/p\rangle$ can be found by averaging over the
statistics of turbulence: $\lambda_1=\lim_{t\to\infty}\langle {\dot p}(t)/p(t)\rangle_u$.
It is useful to consider first the Kraichnan model where $\nabla_j u_i$ in Eq.~\eqref{viscousseparation}
is modeled by white noise ${\hat \sigma}_{ij}$ obeying
$\langle {\hat \sigma}_{ij}(t){\hat
\sigma}_{mn}(t')\rangle=D\delta(t-t')\left[(d+1)\delta_{im}\delta_{jn}-\delta_{ij}\delta_{mn}
-\delta_{in}\delta_{mj}\right]$ where $d$ is the space dimension
\cite{Piterbarg}. Passing to dimensionless time $s \equiv Dt$, one
finds $(D\tau) {\ddot {\bm R}}+{\dot {\bm R}}=\sigma'(s){\bm R}$,
where $\langle
\sigma'_{ij}(s_1)\sigma'_{mn}(s_2)\rangle=\delta(s_1-s_2)\left[(d+1)\delta_{im}\delta_{jn}-\delta_{ij}\delta_{mn}
-\delta_{in}\delta_{mj}\right]$. At $D\tau\ll 1$ one may drop 
$(D\tau) {\ddot {\bm
R}}$. 
The resulting equation is the same as for separation of fluid
particles so $\lambda_1\approx \lambda_1^{fl}$, where $\lambda_1^{fl}\sim D$ is the Lyapunov exponent of fluid particles \cite{review}.
On the other hand, using dimensionless time
$s'=D^{1/3}t/\tau^{2/3}$, one finds ${\ddot {\bm R}}+{\dot {\bm
R}}/(D\tau)^{1/3} =\sigma'(s'){\bm R}$, cf.\ \cite{Horvai}. At
$(D\tau)^{1/3}\gg 1$ one may drop 
${\dot {\bm
R}}/(D\tau)^{1/3}$. Then, by dimensional analysis,
$\lambda_1\sim D^{1/3}/\tau^{2/3}$. We observe that at $\lambda_1\tau\ll 1$ one
can drop the first, inertial, term in ${\ddot {\bm R}}+{\dot {\bm R}}/\tau=
{\hat \sigma}{\bm R}/\tau$, while at $\lambda_1\tau\gg 1$, the friction term can be dropped:
the characteristic time of variations of $\bm R$ is
$\lambda_1^{-1}$ so the ratio of $\ddot {\bm R}$ to $\dot {\bm
R}/\tau$ is estimated as $\lambda_1\tau$. The region of small inertia $\lambda_1\tau\sim D\tau\ll 1$
is separated from the region of large inertia $\lambda_1\tau\sim (D\tau)^{1/3}\gg 1$ by the long
crossover region $D\tau\gtrsim (D\tau)^{1/3}\sim 1$ where $\ddot {\bm R}\sim \dot {\bm
R}/\tau$ and $\lambda_1\tau\sim 1$.
This explains the
numerical results of \cite{Horvai,BecCenciniHillerbrand}.
For example, in $d=2$ one may write
$\lambda_1\tau=[D\tau]^{1/3} \tilde\lambda_1[(D\tau)^{-1/3}]$, where
$\tilde \lambda_1(\epsilon)=\langle {\Re z}\rangle$ for complex
dynamics ${\dot z}=-z^2-\epsilon z + \gamma$ \cite{Piterbarg}.  Here uncorrelated noises
$\gamma_i$ obey $\langle \gamma_1(t)
\gamma_1(t') \rangle = \delta(t-t')$ and $\langle \gamma_2(t)
\gamma_2(t') \rangle = 3\delta(t-t')$.
The crossover region
is described by slow, order unity,
variation of ${\tilde \lambda}_1(\epsilon)$ from ${\tilde \lambda}(1)\approx 0.5$ to $\lambda^0\equiv {\tilde \lambda}_1(0)\approx 2$ \cite{BecCenciniHillerbrand}, and $\lambda_1\approx \lambda^0 D^{1/3}/\tau^{2/3}$ at $(D\tau)^{1/3}\gg 1$.
The time-scale beyond which
$\langle{\dot p}(t)/p(t)\rangle$ relaxes to its steady-state value $\lambda_1$, forgetting the initial
conditions (for $d=2$, the relaxation time of $z$), can be estimated
as $\lambda_1^{-1}$. 
Indeed, $\lambda_1^{-1}$ is the only time-scale both at
$D\tau\ll 1$ and $(D\tau)^{1/3}\gg 1$,
while at $(D\tau)^{1/3}\sim 1$
all coefficients in ${\ddot {\bm R}}+{\dot {\bm
R}}/(D\tau)^{1/3} =\sigma'(s'){\bm R}$ are of order unity so again $\tau^{2/3}/D^{1/3}\sim
\lambda_1^{-1}$ is the only possible time-scale.

Let us now consider the turbulent velocity gradient {\it seen by the particle}, $\nabla_ju_i({\bm
x}(t), t)$ in Eq.~\eqref{viscousseparation}. It  
is determined by fluctuations at the viscous scale,
$\nabla u\sim u_{\eta}/\eta$, while its time variation can be inferred from
\begin{eqnarray}&&
\frac{d}{dt} \bm\nabla \bm u({\bm
x}(t), t) = \left[\left(\partial_t+\bm u\cdot\bm\nabla\right) + \bm w \cdot
\bm\nabla\right] \bm\nabla \bm u.
\label{velder}
\end{eqnarray}
At $\tau\gg t_{\eta}$, the drift derivative $(\bm w \cdot\bm\nabla)\nabla u\sim \nabla u(w/\eta)$
dominates the substantial derivative $\left(\partial_t+\bm u\cdot\bm\nabla\right)\bm\nabla \bm u\sim
\nabla u(u_{\eta}/\eta)$. 
Thus $\nabla_ju_i({\bm x}(t), t)$ varies at time-scale $\eta/w$
-- during this time particle deviates from the carrying flow by the spatial
scale of variations of velocity gradient, $\eta$.

Below we study $\lambda_1$ as a function of the Stokes number ${\rm St}\equiv \lambda_1^{\text{\it turb}}\tau$. 
Here $\lambda_1^{\text{\it turb}}$ is the Lyapunov exponent of fluid
particles in turbulence. K41 dimensional analysis gives $\lambda_1^{\text{\it turb}}\sim \sqrt{\epsilon/\nu}$,
while it does not fix $\lambda_1$ due to the additional time-scale $\tau$. At physically relevant Reynolds numbers
$\lambda_1^{\text{\it turb}}\sim \sqrt{\epsilon/\nu}$ is valid
\cite{BBB}, indicating that $\lambda_1^{\text{\it turb}}$ is determined by weakly intermittent events,
and implying the same for $\lambda_1$ at least for small ${\rm St}$ where $\lambda_1\approx\lambda_1^{\text{\it turb}}$.
We shall assume that $\lambda_1$ is determined by weakly intermittent events for any ${\rm St}$ and use K41 for order of magnitude estimates, justifying this later (e. g. $t_{\eta}$ and $\eta$ will refer to their K41 values).
The domain of applicability of
$\lambda_1\approx\lambda_1^{\text{\it turb}}$ is ${\rm St}\ll 1$ -- at these ${\rm St}$ the drift contribution
in Eq.~\eqref{velder} is negligible, while $\lambda_1\tau\ll 1$ allows to neglect the inertial term
in Eq.~\eqref{viscousseparation}. We now consider ${\rm St}\gg 1$.
The idea of the analysis is to make the
natural assumption $\lambda_1\ll \lambda_1^{\text{\it turb}}$, easily verifiable in the end of the calculation,
and
to use that the timescale $
\lambda_1^{-1}$ of variations of $\bm R$ is much larger than the time-scale of variations
of $\bm \nabla\bm u$ in Eq.~\eqref{viscousseparation}: $\lambda_1^{-1}\gg
[\lambda_1^{\text{\it turb}}]^{-1}\gtrsim \eta/w$. Then the effect of $\nabla \bm u$ on $\bm R$ can
be represented by a white noise.

The first step is the derivation of
the Lyapunov exponent $\lambda_1(a)$ for the auxiliary constant-drift problem
${\ddot R_i}+{\dot R_i}/\tau=R_j\nabla_ju_i(\bm x_{\bm a}(t), t)/\tau$,
where ${\dot {\bm x}_{\bm a}}=\bm u(\bm x_{\bm a}(t), t)+\bm a$. Having in mind the application,
the constant vector $\bm a$ is assumed to have the same characteristic value $\sqrt{\epsilon \tau}$ as
$w$, so that the correlation time $\tau_c$ of $\bm \nabla \bm u(\bm x_{\bm a}(t), t)$ is $\eta/a\sim
1/\lambda_1^{\text{\it turb}}\sqrt{{\rm St}}$.
Averaging Eq.~\eqref{viscousseparation} over time $\Delta t$
satisfying $\tau_c\ll \Delta t\ll \lambda_1^{-1}$, we find $ {\ddot
R_i}+{\dot R_i}/\tau=R_j\overline \nabla_j u_i/\tau$, where
$\overline\nabla_j u_i \equiv \int_t^{t+\Delta t}\nabla_j u_i(\bm
x_{\bm a}(t'), t') \,dt'/\Delta t$. We observe that
$\overline\nabla_j u_i$ is a Gaussian process
with zero mean and
pair correlation which -- due to stationarity, spatial homogeneity
of small-scale turbulence and incompressibility, -- is determined by
$F_{ijmn}({\bm a})\equiv \int \!dt\, \langle \nabla_j u_i(\bm 0, 0)
\nabla_n u_m({\bm q}(t)+{\bm a}t, t)\rangle$. Then
$\overline\nabla_j u_i(t)$ is statistically equivalent
to $\int_t^{t+\Delta t}\sigma_{ij}({\bm a},
t')dt'/\Delta t$, where $\sigma_{ij}({\bm a}, t)$ is a white noise:
\begin{equation}
 \langle\sigma_{ij}({\bm a}, t)\rangle=0,\ \ \ \  \langle \sigma_{ij}({\bm a},t) \sigma_{mn}({\bm a},t') \rangle
=
  \delta(t'-t)F_{ijmn}({\bm a})
\,.\nonumber
\end{equation}
Dropping the auxiliary time-averaging we conclude that $\lambda_1(a)$ can be found
from the anisotropic Kraichnan model ${\ddot {\bm R}}+{\dot {\bm R}}/\tau=\sigma\bm R/\tau$.
For $\eta/a\ll t_{\eta}$ the time variation of $\nabla_ju_i(\bm x_{\bm a}(t), t)$ is determined by
the drift and one can simplify $F_{ijmn}({\bm a}) \approx \int dt \langle \nabla_j
u_i(\bm 0) \nabla_n u_m({\bm a}t)\rangle$.
The arising degeneracy
$a_nF_{ijmn}= \int dt\partial_t\langle \nabla_j u_i(0) u_m({\bm a}t)\rangle=0$ allows to set $\sigma_{i3}\equiv 0$, where we chose
$z-$axis parallel to ${\bm a}$.  As a result, ${\bm
r}\equiv (R_1, R_2)$ satisfies closed equation ${\ddot {\bm
r}}+{\dot {\bm r}}/\tau={\hat \sigma}{\bm r}/\tau$ where ${\hat \sigma}$ is a $2\times
2$ matrix with ${\hat \sigma}_{ij}=\sigma_{ij}$, while ${\dot R_3}=\sigma_{3i}r_i$.
Using $\langle \nabla_{j}u_{i}(\bm 0) \nabla_nu_m({\bm r}) \rangle =
\nabla_j\nabla_n S_{im}({\bm r})/2$, where $S_{ij}({\bm r}) =
\langle [u_i({\bm r})-u_i(\bm 0)] [u_j({\bm r})-u_j(\bm 0)]
\rangle$, one can express $F_{ijmn}$ with the help of second order
structure function of turbulence $S_2(r)=\langle \left(\left[{\bm
u}({\bm r})-{\bm u}(\bm 0)\right]\cdot{\bm r}/r\right)^2\rangle$.
After straightforward calculation one finds that ${\hat \sigma}$ obeys the statistics
of $2d$ Kraichnan model with $D=D(a)= a^{-1} \int_0^{\infty}
S_2(r)/(2r^2)\,dr$. Thus $\lambda_1(a)=D(a)^{1/3}\tilde\lambda_1[(D(a)\tau)^{-1/3}]/\tau^{2/3}$, for $\eta/a\ll t_{\eta}$ or $\sqrt{{\rm St}}\gg 1$. In general, $\lambda_1(a)\tau\sim (D(a)\tau)^{1/3}$ at ${\rm St}\gg 1$.

We now show $\lambda_1\sim \langle \lambda_1(w)\rangle$, where the average is over the single-time
statistics of $w$. If we let $\bm a$ above become slow function of time, then $p(t)$ grows with local
Lyapunov exponent,
$\langle {\dot p}(t)/p(t)\rangle\approx \lambda_1[a(t)]$, 
provided the characteristic time $\lambda_1^{-1}(a)$ of the relaxation of $\langle {\dot p}(t)/p(t)\rangle$ to its local value (see the discussion on Kraichnan model) is much smaller than the characteristic time of variations of $\bm a(t)$.  Next, the drift velocity $\bm w(t)$ is determined by velocity fluctuations with $t_l\gtrsim \tau$ and thus its fixation influences weakly the statistics of velocity gradients determined by fluctuations with time-scale $t_{\eta}\ll \tau$. Thus, $\langle {\dot p}(t)/p(t)\rangle$ for Eq.~\eqref{viscousseparation} can be found first averaging over the gradients at fixed $\bm w(t)$
and then averaging over $\bm w(t)$. The first averaging gives $\lambda_1[w(t)]$ provided
$\lambda_1^{-1}[w(t)]$ is much smaller than the characteristic time $\tau$ of variations of $\bm w(t)$. However,
under the latter condition Kraichnan model gives $\lambda_1[w(t)]=\lambda^0 D^{1/3}[w(t)]/\tau^{2/3}$. Using the expression for $D(a)$ and averaging over $\bm w$, we find
\begin{equation}
\label{lambdafinal} \lambda_1 \approx G,\ \ \text{for\ \
$G\!\equiv\!
   \lambda^0 \big\langle w^{-1/3}\big\rangle
  \left(\!
    \int_0^{\infty}\!\! \frac{S_2(r)dr}{2\tau^2r^2}
  \!\right)^{1/3}\!\!\!\!\gg\! \frac{1}{\tau}$.}
\end{equation}
For $\lambda_1^{-1}[w(t)]\sim \tau$ the above procedure gives the order of magnitude
estimate $\lambda_1\sim \langle \lambda_1(w)\rangle$. Using the results on Kraichnan model,
this gives $\lambda_1\tau\sim 1$ at $G\tau\sim 1$.
Moderate order moments $\langle w^{-1/3}\rangle$ and $\int_{0}^{\infty} S_2(r)dr/r^2$ entering $G$
are expected to be well described by K41 at realistic ${\rm Re}$.
Indeed using for the estimates the multifractal model \cite{Frisch} to analyze $\int_0^{\infty}S_2(r)dr/r^2$
and express the anomalous exponents of $w$ via empirical values of spatial anomalous exponents
\cite{SreeeniFalk}, one finds that
intermittency becomes important at ${\rm Re}$ well above $10^{15}$.
Thus the above use of K41 for order
of magnitude estimates is self-consistent. For $G$ we obtain $G\tau\sim{\rm St}^{1/6}$.
We find the behavior of $\lambda_1$ similar to
Kraichnan model. The region
of weak inertia, $\lambda_1\approx \lambda_1^{\text{\it turb}}\ll 1/\tau$, is separated from
the region of strong inertia $\lambda_1\tau\gg 1$ where Eq.~\eqref{lambdafinal} holds, by the long crossover
${\rm St}\gtrsim {\rm St}^{1/6}\sim 1$ where $\lambda_1\tau\sim 1$
is a slowly varying function of ${\rm St}$ (note $\lambda_1\tau\sim \langle (D(w)\tau)^{1/3}{\tilde \lambda_1}[((D(w)\tau)^{-1/3})]\rangle$ at $\sqrt{{\rm St}}\gg 1)$. The limit
${\rm St}^{1/6}\gg 1$ at $\tau\ll t_L$ means very large ${\rm Re}$ and is of theoretical value mainly.
Summarizing:
\begin{equation}
\lambda_1/\lambda_1^{\text{\it turb}} \sim
  {\rm St}^{-5/6}
\qquad \text{for\ \ ${\rm St}\gtrsim 1$,} \label{answer0}
\end{equation}
The decay of $\lambda_1/\lambda_1^{\text{\it turb}}$ at increasing $\tau$
is faster than in Kraichnan
model because of $D\sim \tau^{-1/2}$.

We now consider the growth of ${\bm R}$ in the inertial range, at
$\eta\lesssim R\ll l_{\tau}$.  In contrast to Richardson's law for
fluid particles $R(t)\sim \epsilon^{1/2}t^{3/2}$ \cite{Frisch,review}, K41 dimensional analysis does
not fix the separation law for inertial particles, due to the
additional time-scale $\tau$.
We shall assume moderate ${\dot {\bm R}}(0)$ not to have mere ballistic motion, e. g. the
analysis below applies to $R(0)\sim \eta$, ${\dot R}(0)\sim \lambda_1\eta$, holding
after separation at $R\ll \eta$. As we will see, $R(t)$ reaches $l_{\tau}$ within $t\sim \tau$, so
to study separation at $R(t)\ll l_{\tau}$ we assume $t\ll \tau$.  The "friction" term ${\dot {\bm R}}/\tau$ in Eq.~\eqref{eq:2p-gen} produces
negligible effect over $t\ll \tau$ and can be omitted.
Also 
$\delta {\bm u}({\bm R})
\approx {\bm u}({\bm q}(t)+{\bm w}t+{\bm R}(t), t)-{\bm u}({\bm q}(t)+{\bm w}t, t)$,
where $\bm w\equiv \bm w(0)$.
The correlation time $t_c(R)$ of $\delta \bm u(R)$ is due to the drift,
$t_c(R)\sim R/w\lesssim t_R$. As we verify later, the time-scale
$\tau_c(R)$ of variations of $R$ obeys $\tau_c(R)\gg t_c(R)$.
Proceeding like in the viscous range,
we introduce Langevin description of $\delta u_i({\bm R})$, substituting it by white noise
$D_{ij}({\bm R})\gamma_j$, where $\langle\gamma_i(t)\gamma_j(t')\rangle=\delta_{ij}\delta(t-t')$.
Here 
$D_{ik}({\bm R})D_{jk}({\bm R}) =\int dt \langle [u_i({\bm R})-u_i(0)]
[u_j({\bm w}t+{\bm R})-u_j({\bm w}t)] \rangle$ to provide the correct dispersion of the time-averaged
$\delta \bm u(\bm R)$ \cite{FH}. We assumed for simplicity
$R/w\ll t_R$ or $(l_{\tau}/R)^{1/3}\gg 1$ (we use K41 as at $R\ll \eta$), so that the time correlations of $\delta \bm u(\bm R)$ are
determined by the drift (cf. to $\eta/a\ll t_{\eta}$
at $R\ll \eta$). 
The above
Kraichnan model for particles is not the same as used usually to model turbulence in the inertial range
\cite{review}: $t_c(R)$ depends on $R$ differently than $t_R$.
Noting 
$D_{ik}({\bm R}) D_{jk}({\bm R})\sim S_2(R)t_c(R)\sim S_2(R)R/w$, we conclude that the dependence on $\epsilon$ and $\tau$ in
${\ddot R_i}=D_{ij}({\bm R}, {\bm w})\gamma_j/\tau$ is via single parameter
$\epsilon^{2/3}/w\tau^2\sim l_{\tau}^{1/3}/\tau^3
$. 
Now dimensional analysis is enough to fix the answer. We find
$\tau_c(R)
\sim\tau (R/l_{\tau})^{1/9}$, so
the applicability condition $\tau_c(R)\gg R/w$ gives
$(R/l_{\tau})^{8/9}\ll 1$, close to just $R\ll l_{\tau}$. At $t\gg \tau_c[R(0)]$
the initial condition is forgotten (we assume explosive separation characteristic of
the inertial range \cite{review}) and $R(t)$ depends only on $t$ and
$l_{\tau}^{1/3}/\tau^3$ giving
\begin{equation}
\label{seplaw} R(t)
\sim l_{\tau}(t/\tau)^{9},\ \ \tau_c[R(0)]\ll t\ll \tau.
\end{equation}
In Kraichnan model, the power-law exponent for dispersion of fluid particles grows indefinitely
as the flow becomes less rough 
(for smooth flow separation is exponential) \cite{review}.
Thus in Eq.~\eqref{seplaw} exponent larger than in Richardson's law
can be attributed to effectively smoother turbulence felt by particles.
As $R(0)\gtrsim\eta$, observability of the power-law entails
$\tau\gg \tau_c[R(0)]\gtrsim \tau_c(\eta)$. This gives
$(l_{\tau}/\eta)^{1/9}\sim {\rm St}^{1/6}\gg 1$, equivalent to the
natural "forgetting" condition $\lambda_1^{-1}\ll \tau$.
At ${\rm St}^{1/6}\sim 1$, the time of forgetting of the initial condition obeys
$\tau_c[R(0)]\sim \tau$ so $R(t)$ at $t\ll \tau$ depends on the details of
initial conditions. Eq. (\ref{seplaw}) then can be used as order of magnitude estimate at $t\sim \tau$
giving $R(\tau)\sim l_{\tau}$. This is expectable - for {\it fluid} particles the time of separation to $l_{\tau}$
is of order $\tau$ and determined by the stage of evolution with $R(t)\sim l_{\tau}$
where fluid and inertial particles behave similarly.

To summarize, we have shown that the difference of the single particle velocity and the local
velocity of the flow grows with inertia as the Lagrangian velocity increment of turbulence at
time $\tau$. While at ${\rm St}\ll 1$ particles disperse like fluid parfticles,
at ${\rm St}\gg 1$ there is a scale $l_{\tau}\sim \eta {\rm St}^{3/2}$
below which particle dispersion obeys laws special for inertial particles and contribution
of not too intermittent events 
can be described by white noise.
The Lyapunov exponent $\lambda_1\sim \lambda_1^{\text{\it turb}}/{\rm St}^{5/6}$ 
is estimated as the average of the Lyapunov exponent for constant drift problem over the drift
velocity, the estimate becoming exact at ${\rm St}^{1/6}\gg 1$.
In the inertial range, $\eta\lesssim R\ll l_{\tau}$, the analogue of Richardson's law is
$R(t)\sim l_{\tau}(t/\tau)^9$. 
The law, observable only at ${\rm St}^{1/6}\gg 1$,
shows property expected at any ${\rm St}\gg 1$: explosive separation to $l_{\tau}$ within $t\sim \tau$,
closer than Richardson's law to exponential separation for smooth flows. The smoothing is by the effective
time-averaging of turbulent velocity difference $\delta \bm u(\bm R)$ driving the separation.
Our treatment can be easily generalized
to incorporate constant gravitational or electric field, where constant drift problem emerges \cite{Grisha,FH} (e.g. for gravity acceleration $\bm g$ one finds ${\bm a}={\bm g}\tau$ and $\lambda_1=\lambda_1[g\tau]$).
The main qualitative result of the work is that collective drift of inertial
particles through the flow makes their relative motion subject to
Langevin description.

We thank J.~Bec, M.~Cencini and R.~Hillerbrand for letting us read
their paper before the publication.  We thank 
K.~Gawedzki and G.~Falkovich for discussions. 

\end{document}